\documentclass[aps,prb,twocolumn,superscriptaddress,showpacs]{revtex4-1}
\usepackage{color}
\usepackage{amsmath}
\usepackage{verbatim}
\usepackage{graphics}
\usepackage{graphicx}
\usepackage{comment}
\usepackage{subfigure}
\usepackage{gensymb}
\usepackage{natbib}
\usepackage{dcolumn}
\usepackage{bm}

\newcommand{\ket}[1]{|#1\rangle}

\newcommand{\vect}[1]{\boldsymbol{#1}}

\begin{document}
\title{Semimetal and Topological Insulator in Perovskite Iridates}
\author{Jean-Michel Carter}
\affiliation{Department of Physics, University of Toronto, Toronto, Ontario M5S 1A7 Canada} 
\author{V. Vijay Shankar} 
\affiliation{Department of Physics, University of Toronto, Toronto, Ontario M5S 1A7 Canada}
\author{M. Ahsan Zeb}
\affiliation{Cavendish Laboratory, Cambridge University, Cambridge, UK}
\author{Hae-Young Kee}
\email{hykee@physics.utoronto.ca}
\affiliation{Department of Physics, University of Toronto, Toronto,
Ontario M5S 1A7 Canada}
\affiliation{Canadian Institute for Advanced Research, Toronto, Ontario,  Canada}

\begin{abstract} 
The two-dimensional layered perovskite
Sr$_2$IrO$_4$ was proposed to be a spin-orbit Mott insulator, where the effect of Hubbard interaction is amplified on a 
narrow $J_{\rm eff}$ = 1/2 band due to strong spin-orbit coupling. On the other hand, the three-dimensional orthorhombic
perovskite (Pbnm) SrIrO$_3$ remains metallic. To understand the physical origin of the metallic state
and possible transitions to insulating phases, we construct a tight-binding model for SrIrO$_3$.
The band structure possesses a line node made of $J_{\rm eff}=1/2$ bands below the Fermi level.
As a consequence, instability towards magnetic ordering is suppressed and the system remains metallic. 
This line node, originating from the underlying crystal structure, turns into a pair of three-dimensional nodal points
on the introduction of a staggered potential or spin-orbit coupling strength between alternating layers.
Increasing this potential beyond a critical strength induces a transition to a strong topological insulator,
followed by another transition to a normal band insulator.  
We propose that materials constructed with alternating Ir- and Rh-oxide layers along the (001)
direction, such as Sr$_2$IrRhO$_6$, are candidates for a strong topological insulator.
\end{abstract}


\maketitle

\section{Introduction}
Recently, there has been intense theoretical and experimental studies on iridium oxides, dubbed iridates,
due to the possibility of realizing spin liquid\cite{Nakatsuji:2006qf, Lawler:2008oq, Chen:2008kl}, topological band insulator\cite{Guo:2009zr, Yang:2010uq,  Wan:2011nx, Jiang:2011cr,Xiao:2011fk}, and topological Mott insulator\cite{Balents,Will:PRB,Yang:2011dq}. %\cite{}
Unlike 3d transition metal oxides, 5d materials are expected to be less correlated
because of extended orbitals. However, due to strong spin-orbit coupling (SOC), the $J_{\rm eff}=1/2$ band
formed out of t$_{2g}$-orbitals is rather narrow, leading to stronger correlation effect of the on-site repulsive Hubbard interaction $U$.
In particular, pyrochlore iridates have been widely studied due to their geometrical frustration,
and novel topological phases have been theoretically proposed.\cite{Guo:2009zr, Wan:2011nx, Yang:2010uq, Balents, Will:PRB, Yang:2011dq}

The perovskite iridates, on the other hand, have attracted less theoretical attention because of the lack of frustration.
Due to this non-frustrated lattice structure, an antiferromagnetic insulator is expected
when the Hubbard $U$ is large compared to the kinetic energy scale as measured by the bandwidth.
It was reported that the SOC and Hubbard interaction are about the same magnitude in 
two-dimensional (2D) layered perovskite Sr$_2$IrO$_4$\cite{Kim:2009bh, Jin:2009dq}, and
a combination of the SOC and Hubbard interaction 
results in a canted antiferromagnetic insulator (AFI). This phenomena was well described by a 
strong Hubbard $U$ limit in Ref. \onlinecite{Jackeli:2009qf}.
While Sr$_2$IrO$_4$ has a canted AFI ground state, three-dimensional (3D) orthorhombic perovskite SrIrO$_3$ remains metallic.\cite{Longo:1971bh, Zhao:2008nx, Moon:2008ly}  

In this paper, we investigate the band structure of orthorhombic perovskite with Pbnm symmetry applicable to SrIrO$_3$
using a tight binding approach.
We show that there is a line node near the Fermi level originating from the underlying lattice structure.
Thus, SrIrO$_3$ is metallic with small Fermi pockets and becomes a perfect 3D semi-metal with nodal dispersion when the SOC becomes stronger. 
When a staggered potential or SOC between alternating layers is introduced, the nodal-line is gapped out leaving 
a pair of 3D nodal points. The system undergoes a transition to a strong topological insulator beyond a critical strength of the potential.
A further increase of the potential turns the system into a trivial band insulator.
The phase diagram is illustrated in Fig. \ref{Fig:PhaseDiagram}, and the nodal band structure along with the mechanism for the transition are presented below.

%%%%%%%%%%%%%%%%%%%%%%%%%%%%%%%%%%%%%%%%%%%%%%%%%%%
\section{Symmetry of lattice and band structure}
The Ruddlesden-Popper series for the strontium iridates Sr$_{n+1}$Ir$_n$O$_{3n+1}$
is a family of materials with Ir atoms surrounded by an octahedral environment of oxygen. 
The octahedral crystal field splits the $d$ electrons to low-energy $t_{2g}$ and high energy e$_{g}$ levels.
The large spin-orbit coupling further splits the t$_{2g}$ electrons into low-lying  $J_{\rm eff} = 3/2$ states and higher-lying $J_{\rm eff} = 1/2$ states.
In these compounds, the oxidation state of the Ir atoms is Ir$^{4+}$ leaving five d-electrons on the outer shell.
When the SOC is larger than Hund's coupling, the $J_{\rm eff}=3/2$ band is completely filled while the $J_{\rm eff}=1/2$ band is half-filled, leading to an expected metallic state based on band theory.

Among the quasi-2D perovskites, Sr$_2$IrO$_4$ is a single layered material where the iridium-oxide layers are separated by strontium atoms\cite{Crawford:1994ys,Huang:1994ve} and the octahedral environment of oxygen around the iridium atoms is rotated around the z-axis in a staggered pattern. $\rm{Sr_{3}Ir_{2}O_{7}}$ is a bilayer version of the single layered material\cite{Subramanian:1994fk}, with opposite rotation of the 
octahedra in the second layer. Both exhibit a canted antiferromagnetic insulating phase due to the effect of the
Hubbard interaction $U$.\cite{Kim:2008ve,Moon:2009ys,Cao:2002uq,Nagai:2007vn}
On the other hand, SrIrO$_{3}$ has an orthorhombic perovskite structure\cite{Longo:1971bh,Zhao:2008nx} 
where the octahedra are rotated around the (110) direction, in addition to the rotation about the z-axis.
This yields a 4-atom unit cell with the space group of Pbnm shown in  Fig. \ref{Fig:3dCrystalStruct}.
We model the different octahedral environment as a rotation around the z-axis by an angle of $\pm \theta$, 
followed by another rotation around the new (110)-direction by an angle of $\pm \varphi$. 
The colors represent Ir atoms with different oxygen environment and 
corresponds to  Blue $= (\theta,\varphi)$, Red $= (-\theta,-\varphi)$, Yellow $= (\theta,-\varphi)$, 
and Green $= (-\theta,\varphi)$.

\begin{figure}[]
\includegraphics[height=1.56in,width=3.4in,angle=-0]{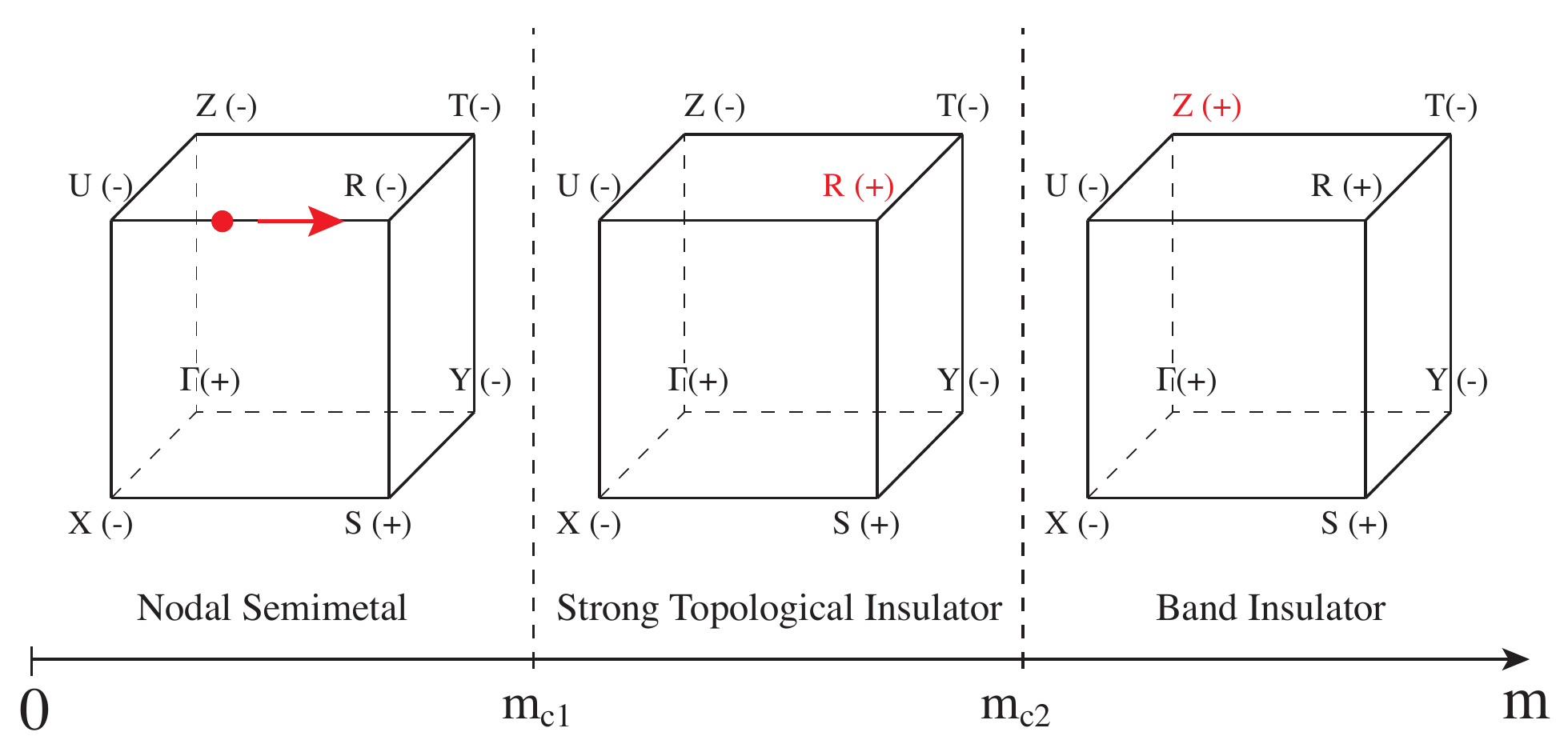}
\caption{[Color online] Phase Diagram as a function of the staggered potential between layers, m. The product of parity eigenvalues for the filled bands
 are denoted by ($\pm$) for each time reversal invariant momentum (TRIM) point. When m is finite, the line node becomes a nodal point located between U and R. This node gets shifted toward R (as indicated by the red arrow) as m increases until it reaches m$_{c1}$, after which a full gap appears. Beyond m$_{c1}$, the system turns into a strong topological insulator with indices $\nu_0;(\nu_1\nu_2\nu_3) = 1;(001)$. At m$_{c2}$, the gap closes at the Z point
and the system becomes a topologically trivial band insulator for $m > m_{c2}$.}
\label{Fig:PhaseDiagram}
\end{figure}

\subsection{Symmetries of Pbnm}
\label{sec:Symmetry}
%***** NEW SECTION *****

The Pbnm structure, in addition to inversion, has the following symmetries: a b-glide, an n-glide and an m reflection. Here, the a-axis is given by the ($\bar{1}$10)-direction, the b-axis by the (110)-direction and the c-axis is concordant with the z-axis.

\begin{figure}[h!]
	\includegraphics[height=2.3in,width=2.66in,angle=-0]{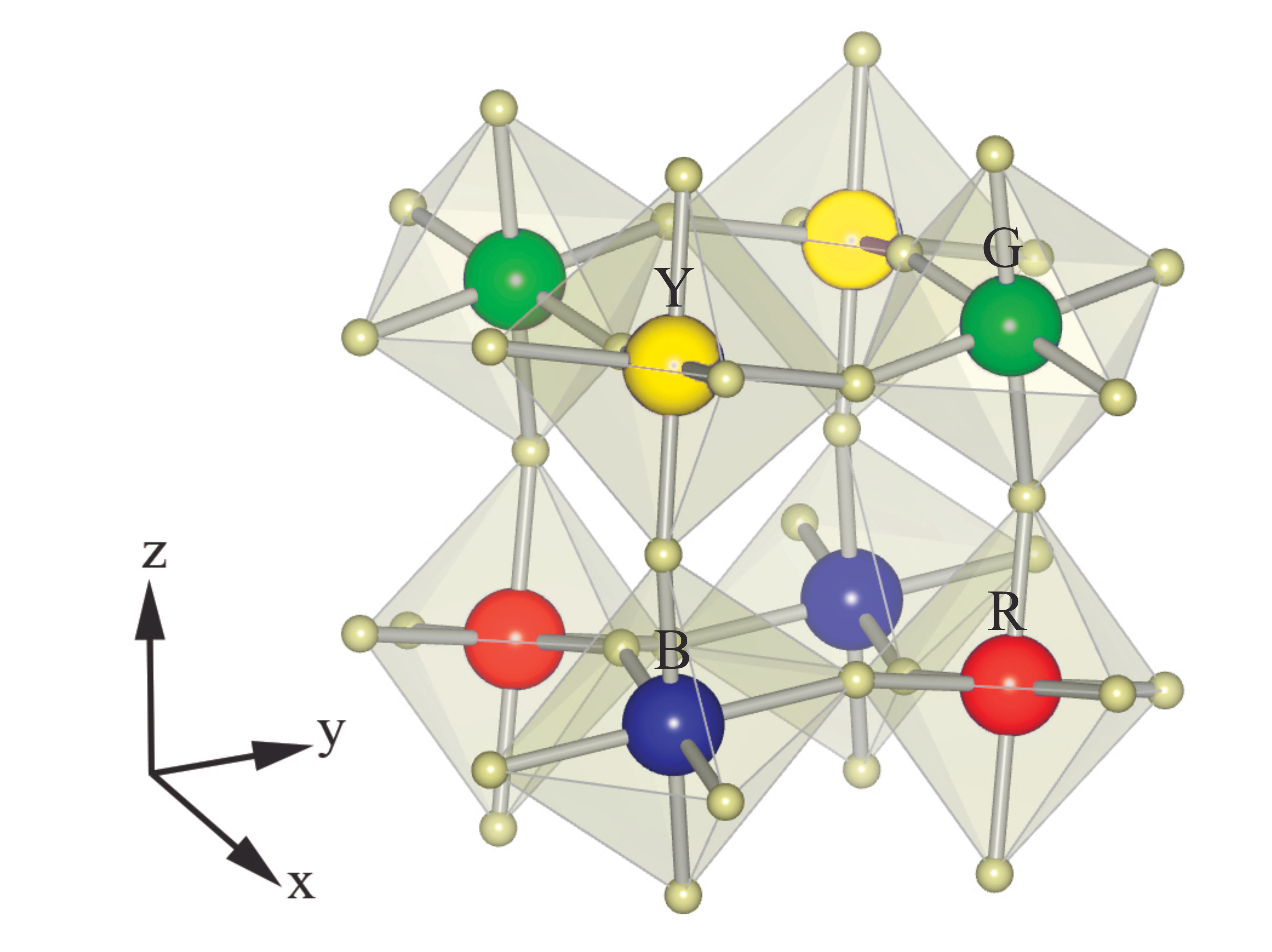}
\caption{[Color online] Crystal structure for SrIrO$_3$ from a side view. The different colors of blue (B), red (R), yellow (Y) and green (G) represent
Ir atoms with different oxygen environment. These 4 Ir atoms form a unit cell}
\label{Fig:3dCrystalStruct}
\end{figure}

The b-glide is represented by a reflection of the a-axis (mirror b-c plane at a = 1/4) followed by a translation along the b-axis.
%The effect of this operation on the lattice degree of freedom, the crystal momentum and the orbital and spin degrees of freedom is given in table \ref{tab:Pib}
The n-glide is represented by a reflection of the b-axis (mirror a-c plane at b = 1/4) followed by a translation along the a-axis and another one along the c-axis.
The m-reflection is represented by a reflection of the c-axis (mirror a-b plane at c = 1/4).
These symmetry operators are denoted by $\hat{\Pi}_b$, $\hat{\Pi}_n$ and $\hat{\Pi}_m$ respectively
The effect of these operations on the Ir atom position, the crystal momentum, the orbitals and the spin is given in Table \ref{tab:Pibnm}.
\begin{table}[ht!]
\caption{Effect of Pbnm symmetry operations on states.}
\label{tab:Pibnm}
\begin{ruledtabular}
%\begin{tabular}{| c || c | c | c | c |}
\begin{tabular}[b]{ c  c  c  c  c }
 				& Lattice 		& ($k_x$,$k_y$,$k_z$) 	& ($d_{xz}$,$d_{yz}$,$d_{xy}$) 	  & spin - ($\uparrow$,$\downarrow$) \\ [0.5ex] % inserts table %heading
\hline
 & & & & \\ [-2ex]
$\hat{\Pi}_b$ 	& $\tau_x$ 		& ($k_y$,$k_x$,$k_z$) 	& ($d_{yz}$,$d_{xz}$,$d_{xy}$) 	  & ($e^{i\pi/4}\downarrow$,$e^{i3\pi/4}\uparrow$)  \\[1ex]
%\hline
$\hat{\Pi}_n$ 	& $\tau_x\nu_x$ & (-$k_y$,-$k_x$,$k_z$) & (-$d_{yz}$,-$d_{xz}$,$d_{xy}$)  & ($e^{-i\pi/4}\downarrow$,-$e^{i\pi/4}\uparrow$)  \\[0.5ex]
%\hline
$\hat{\Pi}_m$ 	& $\nu_x$ 		& ($k_x$,$k_y$,-$k_z$) 	& (-$d_{xz}$,-$d_{yz}$,$d_{xy}$)  & ($i\uparrow$,$-i\downarrow$)  \\
%\hline\hline
\end{tabular}
\end{ruledtabular}
\end{table}
Here the $\tau$ and $\nu$ Pauli matrices represent the two sublattice indices, with $\tau$ for the in-plane sublattice (i.e. blue and red or yellow and green)
and $\nu$ for the sublayer one (i.e. blue and yellow or red and green). %(Blue(Yellow)/Red(Green))

Since the $J=1/2$ basis can be written as %$\ket{J_z=\pm1/2} = \frac{1}{\sqrt{3}}(\ket{d_{yz},\mp s} \pm i\ket{d_{xz},\mp s} \pm \ket{d_{xy},\pm s})$,
\begin{eqnarray}
\ket{J_z=\pm1/2} &=& \frac{1}{\sqrt{3}}(\ket{d_{yz},\mp s} \pm i\ket{d_{xz},\mp s} \pm \ket{d_{xy},\pm s}),  \nonumber
\end{eqnarray}
the orbital and spin (s) transformation can be written for the pseudospin $J = 1/2$ using the $\sigma$ Pauli matrices. Doing so, we can summarize the Pbnm symmetry operations as 
\begin{eqnarray}
\hat{\Pi}_b &=& \frac{i}{\sqrt{2}}(\sigma_x - \sigma_y)\tau_x\vect{k}_b   \nonumber \\
\hat{\Pi}_n &=& \frac{i}{\sqrt{2}}(\sigma_x + \sigma_y)\nu_x\tau_x\vect{k}_n \nonumber \\
\hat{\Pi}_m &=& i\sigma_z\nu_x\vect{k}_m,
\label{eq:Pbnm}
\end{eqnarray}
where $\vect{k}_b$, $\vect{k}_n$ and $\vect{k}_m$ are the operators that act on the crystal momentum as shown in the third column of Table \ref{tab:Pibnm}.

\subsection{First principles calculations}
\label{sec:LDA}
Density functional theory\cite{Hohenberg:1964ys,Kohn:1965rt} calculations including Hubbard $U$ interaction and SOC have been performed using 
the full-potential linearized augmented-plane-wave (FP-LAPW) method as 
implemented in the elk code.\cite{elk} We used the Ceperley-Alder form of the local-density approximation (LDA)\cite{Ceperley:1980vn} as parametrized by Perdew and Zunger\cite{Perdew:1981yq} and used the ``around mean field" (AMF) scheme\cite{Czyzyk:1994kx} for the double-counting correction to the electron-electron interaction. 
To see the effects of reduction or enhancement of the SOC, we simply multiply the corresponding SOC term in the Hamiltonian with a factor $\alpha$; $\alpha=0$ means no SOC and $\alpha=1$ means atomic SOC as implemented by the elk code. 
% if a:  In the FP-LAPW method, the real space is divided into spheres around the atoms (muffin-tins) where atomic orbitals are used as the basis, and interstitials elsewhere where plane waves serve as the basis set. 
% if b:
In the FP-LAPW method, real space is divided into spheres around the atoms (muffin tins) and interstitials elsewhere. 
In our calculations, the muffin-tin radii determined automatically by the elk code are $1.86$, $2.08$, and $1.51$ a.u. for strontium (Sr), iridium (Ir) and oxygen (O), respectively.
% if a:  The basis set we used consists of APW functions with angular momentum $l$ up to $8$ (in the muffin-tin spheres) and plane waves (in the interstitials) with cut-off energy equal to $231.3$ eV. 
% if b: 
The basis set we used consists of APW functions with angular momentum $l$ up to eight and plane waves with cut-off energy equal to $231.3$ eV. 
The number of empty states in the basis set in the second variational step was $10$. Brillouin zone integrations were performed using a $3\times3\times3$ grid, which is equivalent to using $10$ points in the irreducible part of the Brillouin zone. We only used $U$ for the $5d$ orbitals of the Iridium atoms.

\begin{figure}[h!]
	\includegraphics[height=4.2in,width=3.4in,angle=-0]{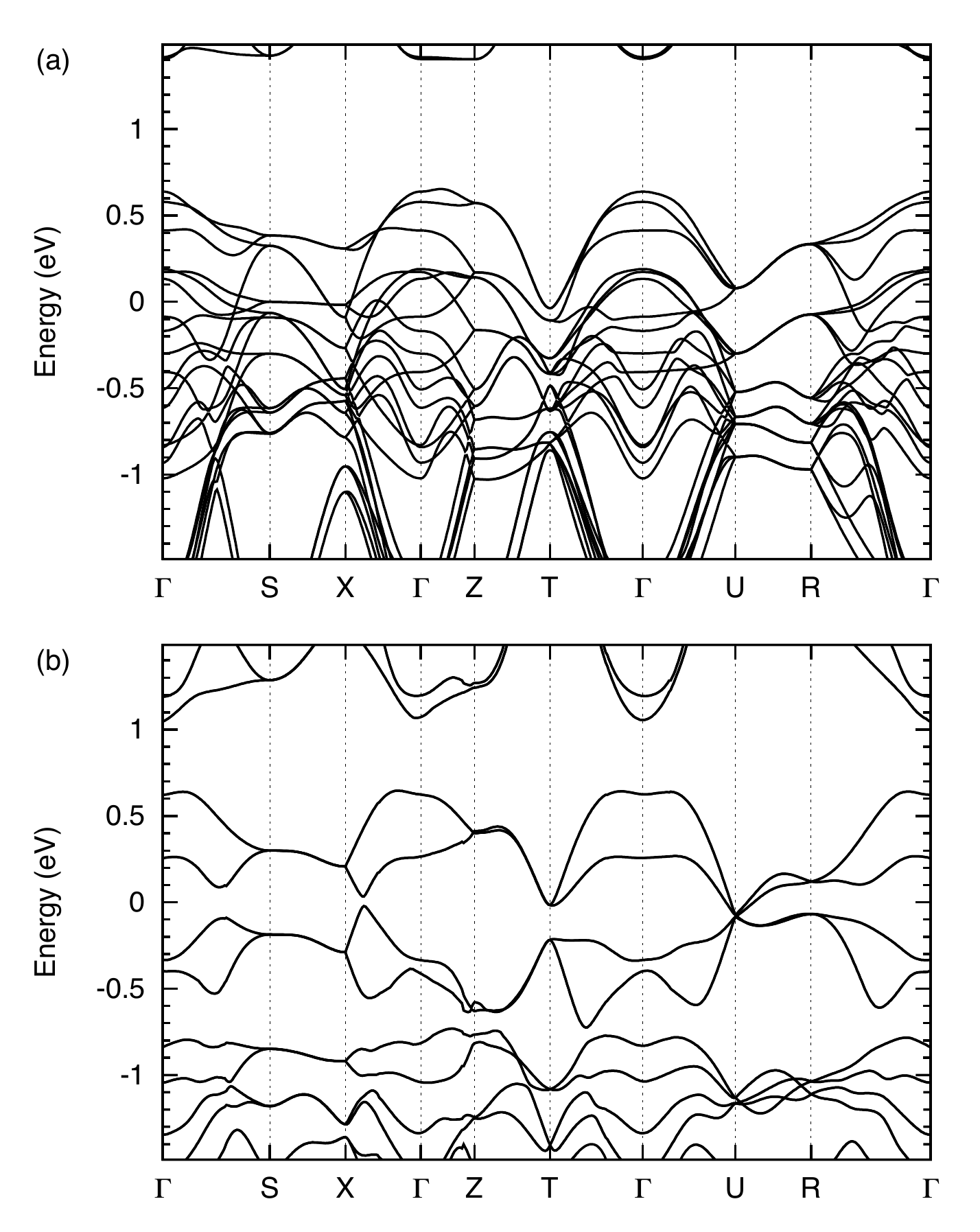}
	\caption{LDA calculation with Hubbard $U = 2$ eV and SOC tuning parameter $\alpha = 0$(a) and $\alpha = 2.0$(b). The band structure remains qualitatively similar from $U = 0$ up to $U = 2$ eV while SOC has a significant effect. See the main text for the comparison to tight binding model.}
	\label{Fig:LDA}
\end{figure}

The computed band structures for $U=2.0$ eV with $\alpha=0$ and $2$ are shown in Fig. \ref{Fig:LDA}. Note that SOC significantly changes the band structure.
In particular near the Fermi energy, the bands are mostly composed of $J_{\rm eff} = 1/2$ states for $\alpha = 2$. 
%The effect of the interaction $U$ is negligible as 
The band structure near the Fermi energy remains qualitatively similar from $U = 0$ to $U=2$ eV, and therefore only the $U = 2$ eV case is shown here.
An important characteristic is the presence of the node near the $U$ point revealing the semi-metallic property of SrIrO$_3$. To understand this feature and possible transition to topological insulating phases, we derive a tight-binding model, described in the next section.
%which reproduce the main features of these first principles calculations very well.

%Fig. \ref{Fig:LDA} shows our results for $U=2.0$ eV and $\alpha=0$ (\ref{Fig:LDA_noSOC}) and for $U=2.0$ eV and $\alpha=2$ (\ref{Fig:LDA_SOC}).
%Our tight binding model described in the next section reproduce the main features of these first principles calculations very well.

\begin{comment}
\begin{figure}[h!]
	\subfigure[]{%Temperature dependences of magnetic susceptibility of the orthorhombic SrIrO3 perovskite at different magnetic fields. Inset: the enlargements in the range of 100–300 K.]{
	\includegraphics[height=2.6in,width=3.8in,angle=-0]{LDA_noSOC_U.pdf}
	\label{Fig:LDA_noSOC}
	}
	\subfigure[]{%Magnetic field dependence of magnetization at 5 K.]{
	\includegraphics[height=2.6in,width=3.8in,angle=-0]{LDA_SOC_U.pdf}
	\label{Fig:LDA_SOC}
	}
	\caption{(a) LDA calculation with $U = 2.0$ eV and $\alpha = 0$ (b) LDA calculation with $U = 2.0$ eV and $\alpha = 2.0$. Note that the effect of U on the bands is very minimal. Our tight-binding model is constructed such that the $J_{eff}=1/2$ bands match closely those of the LDA calculation.}
	\label{Fig:LDA}
\end{figure}
\end{comment}

\section{Tight binding model}
A tight-binding model is constructed, taking into account only the nearest and next-nearest neighbor Ir-Ir direct hoppings
between $J_{\rm eff}=1/2$ orbitals,
using the basis 
$\psi_{\bf k} = (c_{{\bf k B\uparrow}},c_{{\bf k} R \uparrow},c_{{\bf k} Y \uparrow},c_{{\bf k} G \uparrow},
c_{{\bf k} B\downarrow},c_{{\bf k} R\downarrow}, c_{{\bf k} Y\downarrow}, c_{{\bf k} G\downarrow})$ 
where $\{B,R,Y,G\} = \{{\rm Blue, Red, Yellow,Green} \}$ described above, and ($\uparrow$,$\downarrow$) represent $J_{\rm eff}^z=1/2$ and $-1/2$, respectively.
Note that first principles calculations show that all the $J_{\rm eff}$ = 3/2 states are sufficiently below the Fermi energy and the $J_{\rm eff}=1/2$ states occupy bands near the Fermi energy. %\cite{Moon:2008ly}
Thus the current tight binding model of $J_{\rm eff}=1/2$ orbitals is expected to provide relevant bands near the Fermi level.

In such a $J_{\rm eff}=1/2$ basis, the single-particle Hamiltonian is written as
\begin{eqnarray}
	H &=& \lambda_{\bf k} + Re(E^p_{\bf k})\tau_x + Im(E^p_{\bf k})\sigma_z\tau_y + (E^z_{\bf k})\nu_x \nonumber\\
        & &       + Re(E^d_{\bf k}) \nu_x\tau_x  + Im(E^d_{\bf k}) \nu_y\tau_y \nonumber\\
	&	&+ Re(E^{po}_{\bf k})\sigma_y\nu_z\tau_y + Im(E^{po}_{\bf k})\sigma_x\nu_z\tau_y \nonumber\\
	&	&+ Re(E^{zo})\sigma_y\nu_y\tau_z + Im(E^{zo}_{\bf k})\sigma_x\nu_y\tau_z \nonumber\\
	&	&+ Re(E^{do}_{\bf k} )\sigma_y\nu_y\tau_x + Im(E^{do}_{\bf k})\sigma_x\nu_y\tau_x,
\label{eq:Hamil}
\end{eqnarray}
%where the pseudo-spin indices $\tau$, $\nu$, and $\sigma$ represent in-plane sub-lattice, sub-layer, and spin space, respectively.
where the superscripts $p$, $z$, and $d$ represent hopping between Ir atoms
within a layer (e.g., $B$ to $R$), between layers (e.g., $B$ to $Y$), and between next-nearest neighbors (e.g, $B$ to $B$ within a layer or $B$ to $G$ between layers), respectively,
and $o$ denotes hoppings between $J_{\rm eff}^z=1/2$ and $-1/2$.\footnote{These hoppings are non-spin-flip processes in the original orbital and spin basis.}
The coefficients are given by
\begin{eqnarray}
\lambda_{\bf k} & = & \lambda + t_{xy} \cos(\text{kx})\cos(\text{ky}), \nonumber \\
E^p_{\bf k} & =  & (2t - it^{\prime})(\cos (\text{kx})+\cos (\text{ky})), \nonumber \\
E^z_{\bf k}  &= &  2t \cos(\text{kz}),  \;\;\; E^{zo}_{\bf k} = (1-i) t_z^o \cos (\text{kz}), \nonumber\\
E^d_{\bf k}  &= & t_d (\cos(\text{kx})+ \cos(\text{ky}))\cos(\text{kz})  \nonumber\\
             & & + i t_d^{\prime} (\sin(\text{kx}) + \sin(\text{ky}))\sin(\text{kz}), \nonumber\\
E^{po}_{\bf k} & = & \left(t_{1p}^o\cos(\text{kx}) + t_{2p}^o \cos(\text{ky})\right)   \nonumber\\
             && - i \left(t_{2p}^o \cos(\text{kx}) + t_{1p}^o \cos(\text{ky})\right), \nonumber\\
E^{do}_{\bf k} &=& t_d^o(\sin(\text{ky}) + i\sin(\text{kx}))\sin(\text{kz}).
\end{eqnarray}
where $\lambda$ denotes the SOC, $t$ nearest neighbor intra-orbital hoppings, $t_{xy}$ next-nearest neighbor in-plane hopping for $xy$ orbitals, and $t^{\prime}$  the hopping between 1D-orbitals (xz and yz) in the plane. $t_d$ and $t_d^{\prime}$ are originated from intra-orbital diagonal hopping of next-nearest neighbor atoms from different layers (e.g., $B$ to $G$). $t_z^o$ comes from 1D to xy hoppings between two adjacent atoms on different planes whereas $t_{1p}^o$ and $t_{2p}^o$ represent nearest neighbor in-plane hoppings between 1D and xy orbital. Finally, $t_d^o$ arises from out-of-plane next-nearest neighbor 1D to xy orbitals hoppings. 
It is important to note that $t^{\prime}$, $t_d^{\prime}$ , $t_z^o$, $t_{1p}^o$, $t_{2p}^o$ and $t_d^o$ vanish when there is no rotation and/or tilting of the oxygen octahedra.

Note that the symmetry operators of the Pbnm structure described in section \ref{sec:Symmetry} all commute with the Hamiltonian given by Eq. \ref{eq:Hamil}.

The hopping parameters are obtained using the Slater-Koster\cite{Slater:1954tg} method and determined by $(t_{\sigma},t_{\pi},t_{\delta})$ for the nearest neighbor hoppings and $(t_{\sigma}^\prime,t_{\pi}^\prime,t_{\delta}^\prime) = \alpha*(t_{\sigma},t_{\pi},t_{\delta})$ for the next-nearest neighbor hoppings. Thus, the only free parameters are $(t_{\sigma},t_{\pi},t_{\delta})$ and $\alpha$, and the unit of energy is set as $|t_\pi| =1$ throughout the paper. All the other parameters are determined by the rotation and tilting angles of the Pbnm structure. We choose $(t_{\sigma},t_{\pi},t_{\delta}) = (3/2,-1,1/4)$ and $\alpha = 0.4$ for the values of the free parameters to match the LDA band structure shown in Fig. \ref{Fig:LDA}(b). With the angles $(\theta,\varphi) = (11\degree,12\degree)$\cite{Zhao:2008nx}, this yields $(t,t^\prime, t_{xy}, t_z^o, t_d, t_d^\prime, t_{1p}^o, t_{2p}^o, t_d^0) = (-0.6,-0.15,0.13,-0.3,0.03,0.1,0.3,0.06)$ for the tight-binding parameters.

\begin{figure}[h!]
	\subfigure[]{%Temperature dependences of magnetic susceptibility of the orthorhombic SrIrO3 perovskite at different magnetic fields. Inset: the enlargements in the range of 100–300 K.]{
	\includegraphics[height=2.26in,width=3.4in,angle=-0]{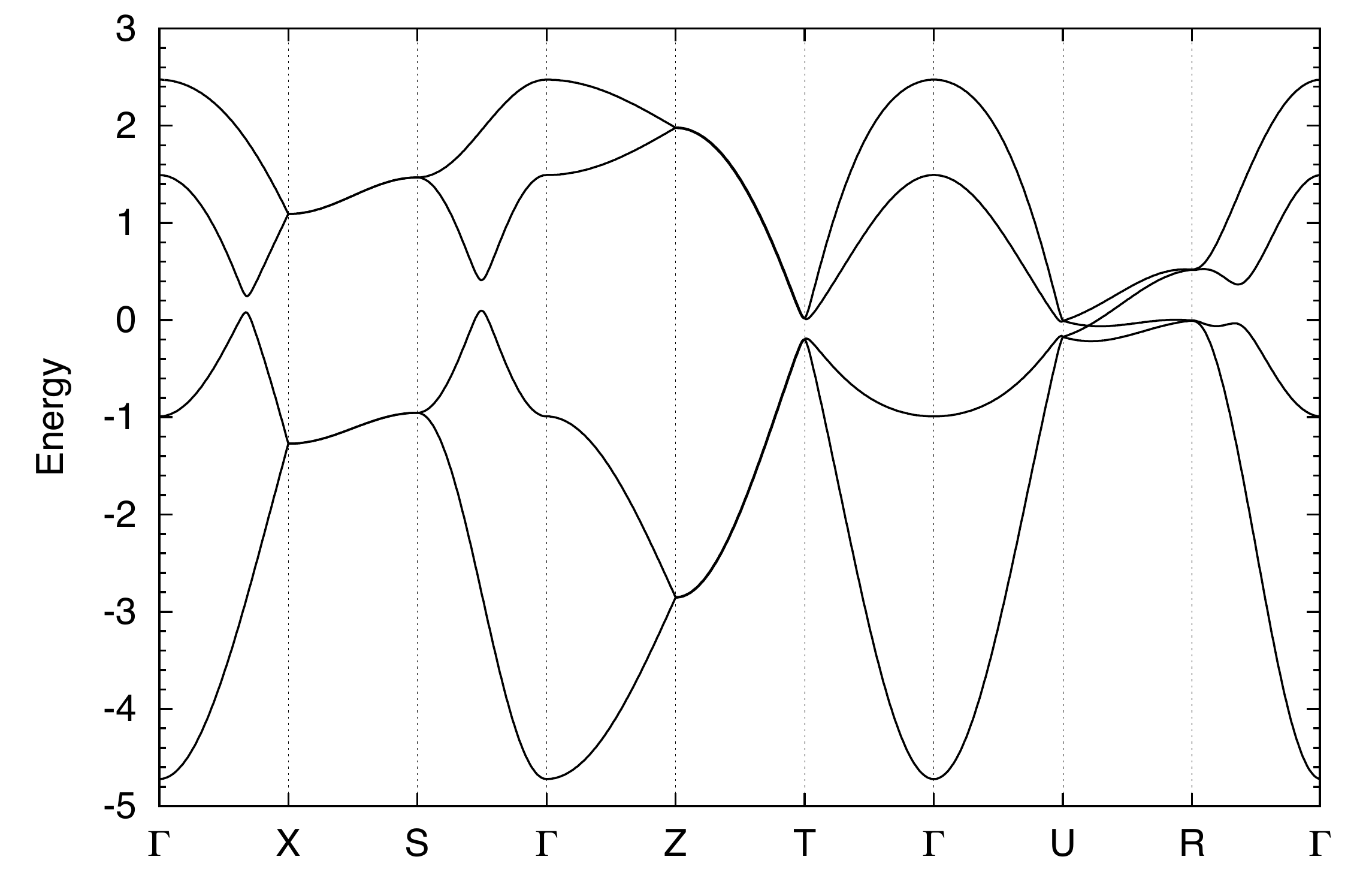}
	\label{Fig:3dBandStruct}
	}
	\subfigure[]{%Magnetic field dependence of magnetization at 5 K.]{
	\includegraphics[height=1.65in,width=1.65in,angle=-0]{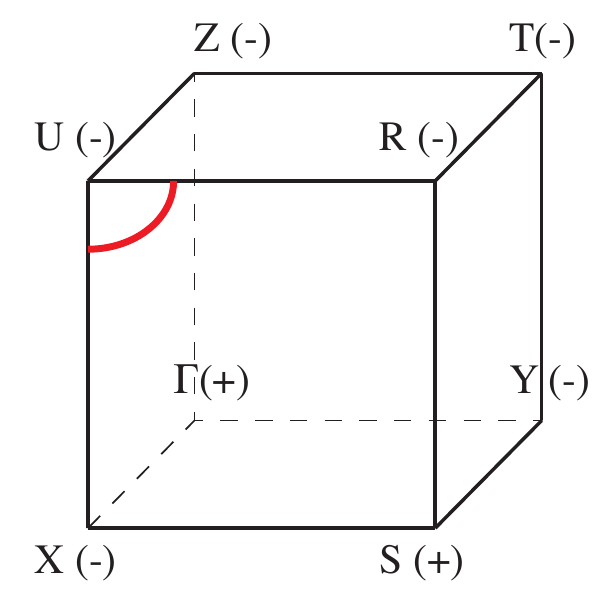}
	\label{Fig:BZParity}
	}
	\caption{[Color online] (a) Underlying band structure of orthorhombic perovskite SrIrO$_3$ obtained by a tight binding model including only
Ir-Ir direct hoppings. Note that the band structure here is qualitatively similar the the LDA calculation of Fig. \ref{Fig:LDA}, especially near the Fermi surface. (b) Brillouin zone with all TRIM points labeled along with the product
of the parity eigenvalues of the filled bands. The red line in the XURS plane corresponds to the line node in the system.} %(*** Split that figure into 2***)
	\label{Fig:TBband}
\end{figure}

The band structure is shown in Fig. \ref{Fig:3dBandStruct}. While the detailed dispersion is sensitive to the rotation and tilting angles, and to the choice
of a parameter set, the robust feature is a line node near the U point in the Brillouin zone, consistent with first-principles calculations. %\cite{Zeb}.
This line node, which is underneath the Fermi level as shown in Fig. \ref{Fig:3dBandStruct}, originates from the reflection symmetry of the crystal structure
at the $z = \frac{1}{4}$ and $z = \frac{3}{4}$ planes present in the Pbnm space group ($\hat{\Pi}_m$).
Without $t_d^o$, there is a node that is 8-fold degenerate at the U $ = (\frac{\pi}{2},-\frac{\pi}{2},\frac{\pi}{2})$ point.
With a finite $t_d^o$, a gap opens at the U point, and the nodes are pushed away from the U point in the XURS plane.
It was found that an elliptical line node,
shown as the red line in Fig. \ref{Fig:BZParity}, appears in the XURS plane.
The line node obtained by expanding the dispersion near the U point up to linear terms is given by
\begin{equation}
	(t_d^o)^2 = (t_{1p}^o + t_{2p}^o)^2 p_+^2 + ((t_z^o)^2 + 2t^2)p_z^2
\end{equation}
where, $p_+$ and $p_z$ are defined such that $k_x = \frac{\pi}{2} + p_+ + p_-$, $k_y = -\frac{\pi}{2} + p_+ - p_-$ and $k_z = \frac{\pi}{2} + p_z$.
The plane of the ellipse appears at $p_- = 0$.
\footnote{For this analysis, we deliberately assumed that $t_{xy} = 0$, but such a term will not lift this line node, but will simply give it some dispersion.  A detailed discussion of line node in various cases can be found in A. A. Burkov, M. D. Hook, and L. Balents, Phys. Rev. B {\bf 84}, 235126 (2011)}%A. A. Burkov {\it et al.}, arXiv:1110.1089 (2011)}

Due to this robust feature of nodes originated from the combination of SOC and Pbnm symmetry of the lattice,
SrIrO$_3$ has small Fermi pockets with small density of states near the Fermi energy.
As a consequence, it is expected that instability toward any  magnetic ordering is suppressed. 
Indeed a larger value of the local Hubbard interaction is required for a canted antiferromagnetic ordering, 
when compared to the quasi-2D perovskite iridates.
%While the Hubbard interaction does not lead to a broken symmetry, it flattens the band and makes the band masses heavier.\cite{Zeb}
As shown in Fig. \ref{Fig:LDA}, the system remains metallic for $U$ up to $2$ eV.
Since the line node is protected by lattice symmetries and is not affected by moderate values of Hubbard interaction,
perturbations that break the sub-layer reflection symmetry ($\hat{\Pi}_m$) at $z=\frac{1}{4}$ and $\frac{3}{4}$ planes are required to open a gap along the U to R line and turn the system into an insulating phase.
Below, we discuss one such perturbation and analyze the resulting phases.

\section{Topological insulator}
%In order to break the reflection symmetry at $z=\frac{1}{4}$ and $z=\frac{3}{4}$ planes,
The mirror symmetry of the Pbnm space group ($\hat{\Pi}_m$) commutes with every term of the Hamiltonian.
In order to break such symmetry, we introduce a staggered sublayer potential.
Such a staggered chemical potential is added to the Hamiltonian in Eq. \ref{eq:Hamil} as a mass term of the form m$\nu_z$, which anticommutes with $\hat{\Pi}_m$, and hence breaks this symmetry.
This term can also arise when the SOC alternates between the layers ($B$-$R$ vs $Y$-$G$ planes along $z$-axis as shown in Fig. 2).
As the mass term is introduced, the degeneracy of the line node is lifted  except for a pair of points
along the U to R line which gets shifted towards the R point as shown by the red arrow in Fig. \ref{Fig:PhaseDiagram}.
Along this line, only two terms of the Hamiltonian are non-zero, only one of which commutes with m$\nu_z$, having the effect of pushing the node instead of lifting it completely. % (**** New sentence ****).
At m = m$_{c1}$, the node reaches the R point, and the system changes into an insulating phase  for $m > m_{c1}$
as shown in Fig. \ref{Fig:PhaseDiagram}.
The insulating phase, above this critical value of the mass term, is classified as a strong topological insulator characterized by the Z$_2$ topological indices.\cite{KaneMeleZ2, FuKaneInversion, TKNN, Hasan_Kane_Review, Qi_Zhang_review}
The product of parity eigenvalues of the filled bands at each TRIM points is shown in Fig. \ref{Fig:PhaseDiagram}, and the topological indices for this phase are $\nu_0;(\nu_1\nu_2\nu_3) = 1;(001)$.
Notice that the product of the parity eigenvalues at the R point has flipped sign at m$_{c1}$.

Further increase of the mass term leads to another transition at m$_{c2}$ where the bands invert at the Z point,
and the product of parity eigenvalues switches sign accordingly as shown in Fig. \ref{Fig:PhaseDiagram}.
Beyond m$_{c2}$, the system becomes a trivial band insulator.
For the set of parameters we have chosen, the values of m$_{c1}$ and m$_{c2}$ are found to be approximately m$_{c1}$ = 0.27 and m$_{c2}$ = 2.4,
forming a large window of strong topological insulating phase.

\begin{figure}[h!]
\includegraphics[height=2.5in,width=3.4in,angle=-0]{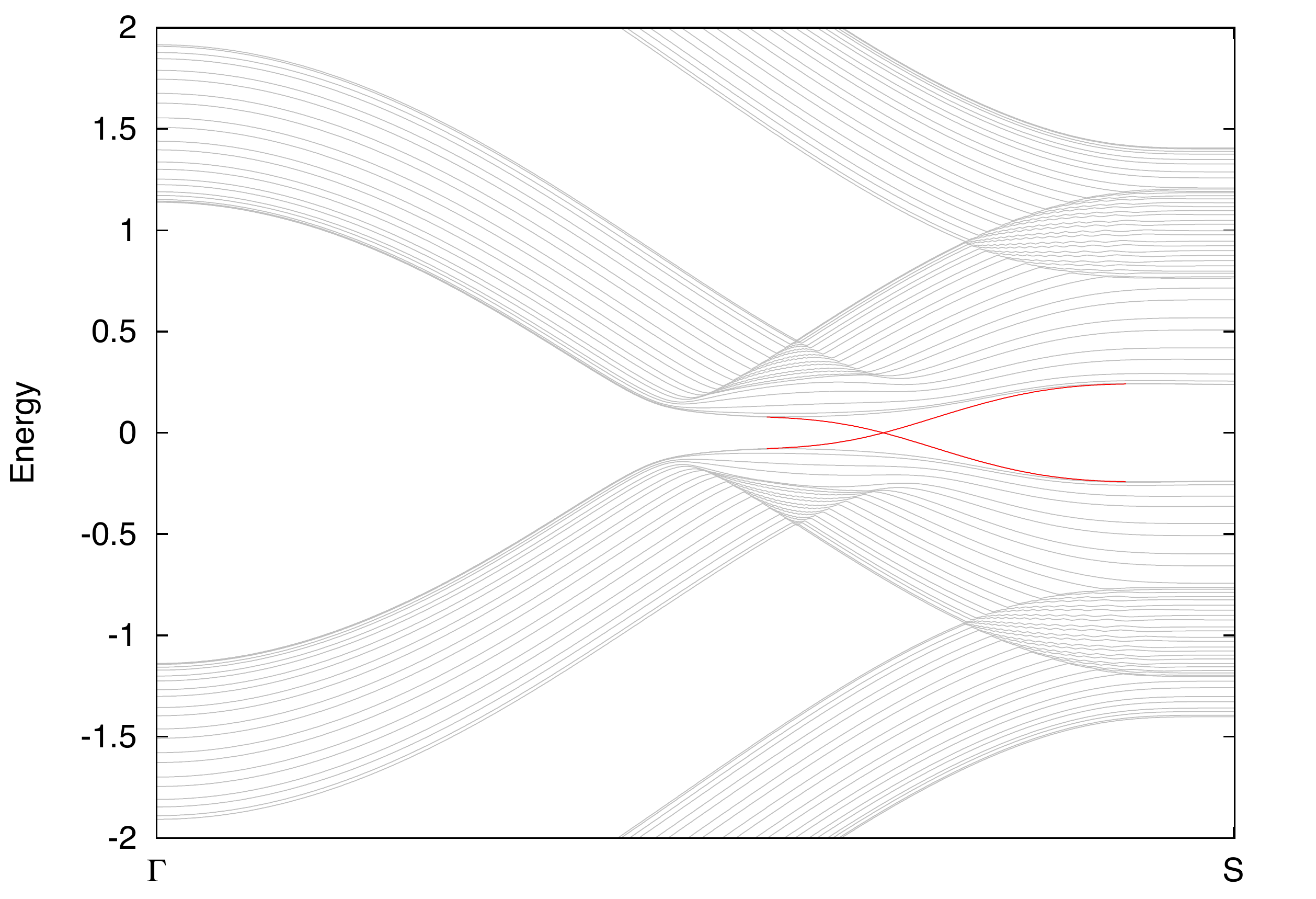}
\caption{[Color online] The edge state calculation with a finite slab of material having periodic boundary conditions along the x and y directions and open boundary condition along the z direction. From the Fu-Kane criteria, 
we expect to see an odd number of crossing of the edge state between $\Gamma$ and any of the other three TRIM points in the surface Brillouin zone. Here, we compute the band structure of such a system between $\Gamma$ = $(0,0)$ and $S = (\pi,0)$ for $m=0.5$.}
\label{Fig:EdgeState}
\end{figure}

To confirm the above Z$_2$ analysis, we compute the surface state of a finite slab to
verify the presence of protected edge states.\cite{KaneMeleZ2, Hasan_Kane_Review}
The band structure calculation using periodic boundary condition along the x and y axis
and open boundary conditions along the z axis is shown in Fig. \ref{Fig:EdgeState}.
The red lines represents the surface states and the calculation was done using a Blue-Red layer as one edge, and a Yellow-Green layer as the other edge.
Because of the sign difference of the mass term on these two layers, the surface states coming from these two edges have opposite velocity.

%\phantom{a}

%\phantom{a}
\vspace{15pt}
\section{Discussion and summary}
Based on the results of band structure obtained for SrIrO$_3$ with Pbnm space group symmetry, a staggered potential in alternating layers
can turn the semi-metal state into a strong topological insulator.
Such a potential can be obtained if the Ir atoms in half of the layers (e.g., Y-G layer) are replaced by Rh atoms
with a Rh$^{4+}$ electronic configuration, leaving 5 d-electrons in the outer 4d shell.
Rh, being a lighter element, has a smaller atomic SOC than Ir.
Hence, adding Rh atoms in this way creates a staggered SOC along the c axis.
We predict that tailoring a superlattice of Sr$_2$IrRhO$_6$ will lift the nodal line present in SrIrO$_3$ and may produce
a strong topological insulator in perovskite iridates.
Sr$_2$IrCoO$_6$ is another candidate as Co$^{4+}$ in 3d shell leads to a similar effect of staggered SOC, but due to a smaller
strength of the SOC in Co, stronger $U$ may be necessary to generate a strong topological insulator.
The band structure of the proposed materials and the effect of the combination of SOC and Hubbard interaction will be presented elsewhere.\cite{Zeb}

Material realization of topological insulators has so far been limited to
narrow band gap semiconductors based on Hg or Bi, where s and p orbitals are
involved. In addition to a recent theoretical proposal, where perovskite heterostructures
of transition metal oxides grown along the (111) direction could be tuned to
topological insulators\cite{Xiao:2011fk},
our results further encourage the experimental search for topological phases via synthesis
and characterization of transition metal oxide materials. In particular, the strong SOC in
Ir makes iridates good candidates.

In summary, we have shown that the underlying band structure of SrIrO$_3$ possesses a line node near the Fermi surface due to the crystal symmetry.
Such a line node can be lifted by breaking the sublayer reflection symmetry and a transition to a strong topological insulator can be achieved.
We suggest that superlattices of Sr$_2$IrRhO$_6$ or Sr$_2$IrCoO$_6$ tailored in the direction suggested in this paper
may be good candidates to achieve a strong topological insulator.

{\it Acknowledgement} --
The work is supported by NSERC of Canada and Canada Research Chair.
JMC and VVS would like to thank Jeffrey G. Rau for useful discussions.

\bibliography{BibSrIrO}

\end{document}